\documentclass{Interspeech2024}

\usepackage{authblk}
\usepackage{multirow}
\usepackage{multicol}
\usepackage[table]{xcolor}
\usepackage{graphicx}
\usepackage{subfigure}
\usepackage{subcaption}
\usepackage{tabularx}
\usepackage{verbatim}

\begin{document}

\interspeechcameraready


\title{When Vision Models Meet Parameter Efficient Look-Aside Adapters\\Without Large-Scale Audio Pretraining}


\name[affiliation={1*}]{Juan}{Yeo}
\name[affiliation={1*}]{Jinkwan}{Jang}
\name[affiliation={1}]{Kyubyung}{Chae}
\name[affiliation={2}]{Seongkyu}{Mun}
\name[affiliation={1}]{Taesup}{Kim}


\address{
  $^1$Graduate School of Data Science, Seoul National University\\
  $^2$Samsung Research}

\email{\{juanyeo, jkjang22, kyubyung.chae, taesup.kim\}@snu.ac.kr, skmoon777@gmail.com}


\maketitle

\renewcommand{\thefootnote}{\fnsymbol{footnote}} 
\footnotetext[1]{Equal contribution.}

\begin{abstract}

    Recent studies show that pretrained vision models can boost performance in audio downstream tasks. To enhance the performance further, an additional pretraining stage with large-scale audio data is typically required to infuse audio-specific knowledge into the vision model. However, such approaches require extensive audio data and a carefully designed objective function. In this work, we propose bypassing the pretraining stage by directly fine-tuning the vision model with our Look-Aside Adapter (LoAA) designed for efficient audio understanding. Audio spectrum data is represented across two heterogeneous dimensions—time and frequency—and we refine adapters to facilitate interactions between tokens across these dimensions. Our experiments demonstrate that our adapters allow vision models to reach or surpass the performance of pretrained audio models in various audio and speech tasks, offering a resource-efficient and effective solution for leveraging vision models in audio applications. 

\end{abstract}

\newcommand\blfootnote[1]{%
  \begingroup
  \renewcommand\thefootnote{}\footnote{#1}%
  \addtocounter{footnote}{-1}%
  \endgroup
}


\vspace{2em}

\section{Introduction}
    In the era of transformers \cite{vaswani17-ATTN}, self-supervised learning \cite{caron21-DINO, chen21-SimSiam} is revolutionizing all domains, including computer vision (CV) and natural language processing (NLP). The paradigm of pretraining followed by fine-tuning has been widely embraced. However, compared to CV and NLP, the audio domain often encounters the challenge of relatively small datasets for large-scale pretraining. Despite the emergence of large-scale audio classification datasets such as AudioSet \cite{gemmeke17-AudioSet} and EPIC-SOUNDS \cite{huh23-EPIC}, there remains a pressing need for more extensive data to support the exponential scaling of transformer-based models.
    
    An approach to audio classification involves initially leveraging pretrained weights from the ImageNet dataset \cite{deng09-IN} in the current research paradigm.
    This strategy has shown superior performance over models with randomly initialized weights, underscoring the potential of cross-modality transfer learning and the efficiency of attention-based models in processing audio datasets, which often suffer from limited resources \cite{Parlanisamy20-RCN, Gong21-AST}.
    
    Subsequently, methods for audio transfer learning, such as the Self-Supervised Audio Spectrogram Transformer (SSAST) and Audio-MAE, require large-scale audio pretraining to acquire domain-specific knowledge \cite{Kong20-PANN, Gong22-SSAST, huang22-AMAE}. Furthermore, an additional fine-tuning stage is required for downstream tasks. If the downstream dataset is as large as the pretraining dataset, total training costs would significantly increase.
    Thus, it would be beneficial to employ a methodology that is parameter efficient and allows for a straightforward adaptation to downstream tasks without dividing the training procedure into multiple stages.
    For example, the concept of convolutional bypasses (Convpass) has been introduced as efficient adaptation modules for Vision Transformers in computer vision research \cite{jie22-CP}.
    
    
    \begin{figure}[t]
      \centering
      \includegraphics[width=\linewidth]{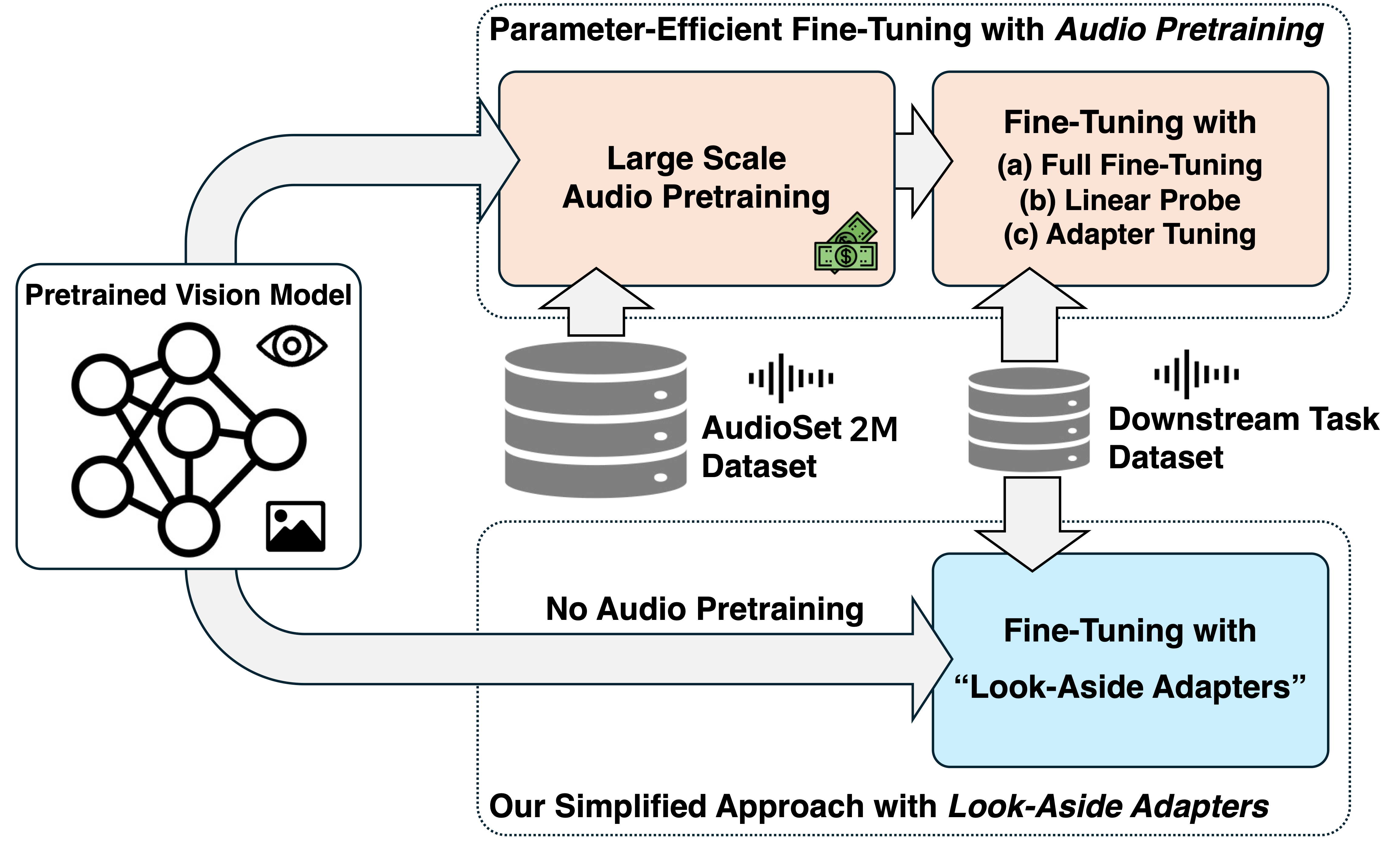}
      \caption{An illustration of our simplified approach for audio classification. Our newly proposed Parameter Efficient Fine-Tuning (PEFT) paradigm for audio classification is a direct adaptation to downstream tasks in a singular stage. This approach even outperforms the current paradigm, which involves pretraining with large-scale audio datasets such as AudioSet-2M, as evidenced by its performance on the EPIC-SOUNDS dataset.}
      \label{fig:PEFTwithLoAA}
    \end{figure}
    
    With a focus on parameter efficiency with simple procedures and adaptation to the specific properties of audio data, we considered the incorporation of audio-friendly adapters alongside image models.
    We propose a new parameter efficient paradigm for audio classification, featuring the Look-Aside Adapter (LoAA).
    This audio friendly adapter is designed to aid ImageNet pretrained vision models in facilitating interactions between tokens across both time or frequency dimensions for audio data.
    Because unlike image data, which contains only spatial information, audio data—when converted to a Mel-spectrogram—encompasses two distinct domains: time and frequency.

    Our evaluation spans various audio and speech datasets, including EPIC-SOUNDS, ESC-50 \cite{piczak15-ESC}, and Speech Commands V2 \cite{warden18-GSC}.
    It is especially notable that our framework for audio classification surpasses the performance of large-scale pretrained models, such as SSAST, on the EPIC-SOUNDS dataset. Note that this performance has been achieved without any large-scale audio pretraining.
    
    \begin{figure}[ht!]
      \centering
      \includegraphics[width=\linewidth]{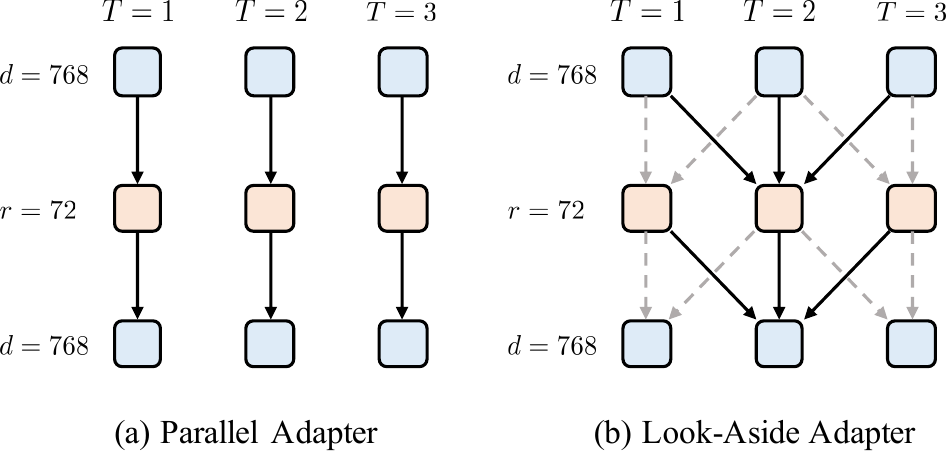}
      \caption{Graphical illustration of adapter module pipelines of conventional parallel adapter (left) and our look-aside adapter (right). $d$: input token dimension; $r$: bottleneck dimension; $T$: time sequence of each token.}
      \label{fig:interactions}
    \end{figure}

\section{PEFT with Look-Aside Adapter}
\subsection{Parameter-Efficient Fine-Tuning (PEFT)}
    Parameter-efficient fine-tuning (PEFT) aims to adapt a pretrained model to unseen domains and downstream tasks by updating only a small number of parameters. For example, adapter tuning \cite{rebuffi2017learning, houlsby19-PETL-NLP} inserts small trainable modules (adapters) to each layer of a frozen pretrained model. Additionally, to enhance efficiency, LoRA \cite{hu22-LoRA} injects low-rank matrices into projection layers within attention blocks. Another approach involves prompt tuning \cite{lester21-PEPT} and prefix tuning \cite{li21-prefixtuning}, which prepends learnable prompts to input or activations. These approaches demonstrated comparable performance to full fine-tuning while updating only a fraction of the model parameters. Our work leverages the PEFT to fine-tune image models for audio downstream tasks with less than 10\% of total parameters updated.

\subsection{PEFT for Audio based on a Frozen Image Model}
    Recent works \cite{Gong21-AST, gong2023contrastive} demonstrated that using ImageNet pretrained weights for initialization improves audio model performance even when they are trained on large-scale audio datasets afterward. This suggests that pre-trained knowledge about image data can be transferred into the audio domain. Still, the performance of a model pre-trained solely on an image dataset \cite{Gong21-AST} is inferior to that of a model pretrained on audio datasets \cite{huang22-AMAE}. This is because, although the audio spectrogram is similar to 2D images, they have different characteristics. Each dimension in image data contains spatial information. However, the audio data is heterogeneous, represented by time and frequency dimension. Most recent works mitigate this gap between modalities by large-scale pretraining on audio datasets \cite{huang22-AMAE, Baade2022MAEASTMA, chong2023masked}. In the training process, the model learns how to distinguish time and frequency dimensions naturally.
    Since this large-scale pretraining is expensive in terms of both computation and memory cost, we propose applying PEFT to image models for multiple audio downstream tasks. 
    
    The PEFT method we use in this work is parallel Houlsby adapter \cite{houlsby19-PETL-NLP}, a simple approach in PEFT that attaches small feed-forward networks next to transformer layers. 
    The adapter module gets input $h$ from block $B(\cdot)$, which can be either an attention (Attn) block or a feed-forward network (FFN) block. The initial linear layer $W_{\downarrow} \in \mathbb{R}^{d\times r}$ projects it to a lower-dimensional space defined by bottleneck dimension $r (r < d)$. The low dimension input then goes through nonlinear activation function $f(\cdot)$ (e.g., GELU) and second linear layer $W_{\uparrow} \in \mathbb{R}^{r\times d}$ that projects it back to the original input dimension $d$, as shown in Figure 2(a). The final output is computed by adding block output and the adapter module output:\begin{equation}
        h \gets B(h) + f(hW_{\downarrow})W_{\uparrow}
    \end{equation}

    The audio input $h$ holds information of three domains: time, frequency, and token dimension. The parallel adapter, employing linear layers, facilitates interactions only within the token dimensional axis, as shown in Figure 2(a). Consequently, in a PEFT scenario with a frozen image model, the interaction between tokens relies entirely on the pretrained attention layers. In other words, employing previous PEFT methods does not directly affect how tokens interact across time and frequency axes. Instead, these methods project each token to an appropriate dimensional space and hope image-based pretrained attention layers perform well. To tackle this issue, we propose a Look-Aside Adapter, which enables inter-token interaction within the adapter module through a simple modification to existing methods.

 \begin{table}[t!]
      \caption{Top-1 accuracy when putting adapters alongside with attention block or feedforward network block for each kernel and the number of parameters on EPIC-SOUNDS dataset. Best results are highlighted in \textbf{bold}, and second best results are \underline{underlined}.}
      \label{attn-ffn-epic-sounds}
      \centering
      \begin{tabular}{lllll}
      \hline
        \toprule
        \multirow{2}{*}{\textbf{\# Param (Ratio)}} & \multicolumn{4}{c}{\textbf{Kernel Shape}} \\ 
                        & \multicolumn{1}{c}{(1,1)}    & \multicolumn{1}{c}{F(3,1)}    & \multicolumn{1}{c}{T(1,3)}    & \multicolumn{1}{c}{(3,3)}   \\ \midrule
        \multicolumn{5}{l}{\textbf{Audio Pretrained Model~(SSAST):} 53.75} \\ \midrule
        \multicolumn{4}{l}{\textbf{ATTN Block}} \\
         1.035M~(1.2\%)  & 53.38    & \textbf{53.79}     & \underline{53.43}     & 53.20   \\
         2.026M~(2.3\%)  & 53.33    & \underline{54.44}     & \textbf{54.52}     & 53.64   \\
         4.017M~(4.6\%)  & 53.42    & \underline{53.86}     & \textbf{54.08}     & 53.44   \\ \midrule
        
        \multicolumn{4}{l}{\textbf{FFN Block}} \\
         1.035M~(1.2\%)  & 53.04    & \textbf{53.53}     & \underline{53.42}     & 53.03   \\
         2.026M~(2.3\%)  & 53.42    & \underline{53.73}     & \textbf{54.50}     & 53.30   \\
         4.017M~(4.6\%)  & 53.30    & \underline{53.73}     & \textbf{54.01}     & 53.24   \\ 
        \bottomrule
      \end{tabular}
      \label{table:kernel}
    \end{table}

\begin{table*}[t!]
  \caption{Performance comparison of PEFT with LoAAs and current paradigm on various audio and speech datasets. T: Time, (1,3) kernel; F: Frequency, (3,1) kernel. AS, LS, and IN in the PT-Data column correspond to AudioSet \cite{gemmeke17-AudioSet}, LibriSpeech \cite{Panayotov15-Librispeech}, and ImageNet \cite{deng09-IN}, respectively. The best results are highlighted in \textbf{bold}, and the second best results are in \underline{underlined}.}
  \label{maintable}
  \centering
  \begin{tabular}{lllllcccc}
    \toprule
    \multicolumn{4}{r}{\multirow{2}{*}{}} & \multicolumn{3}{c}{\textbf{Audio Dataset}}                                                                    & \multicolumn{1}{c}{\textbf{Speech Dataset}}                    \\
    \multicolumn{4}{r}{}                         & \multicolumn{2}{c}{\textbf{EPIC}}    & \multicolumn{1}{c}{\textbf{ESC-50}}  & \multicolumn{1}{c}{\textbf{SPC-2}}        \\
    \textbf{Model} & \textbf{Fine-Tuning Method} & \textbf{\# Param} & \textbf{PT-Data} & \multicolumn{1}{c}{\textbf{Top-1}}  & \multicolumn{1}{c}{\textbf{mAP}}  & \multicolumn{1}{c}{\textbf{mAP}} & \multicolumn{1}{c}{\textbf{mAP}}  \\ \midrule\midrule
    \multicolumn{4}{l}{\textbf{Image Pretrained} $\rightarrow$ \textbf{Audio Pretrained}}\\ \midrule
    MAE-AST~\cite{Baade2022MAEASTMA}   & Full FT               & 100\%    &  AS+LS & -              & -           & 90.0 & 97.9 \\
    SSAST~\cite{Gong22-SSAST}     & Full FT               & 100\%    &  AS+LS & 53.75          & 0.237            & 88.8 & 98.0       \\ \midrule \midrule
    \multicolumn{4}{l}{\textbf{Image Pretrained Only}} \\ \midrule
    AST~\cite{Gong21-AST}        & Full FT               & 100\%    &  IN    & 52.22          & 0.222            & \underline{87.9} & \textbf{97.8}   \\
              & Linear Probe          & 0.04\%   &  IN    & 38.95          & 0.136            & 65.6    & 42.1 \\
     & Parallel Adapter            & 5\%      &  IN    & 53.40          & 0.227            & 87.0   & 96.7 \vspace{0.1em } \\ 
     \cline{2-8}
              & Ours - Attn(T)        & 2\%      &  IN    & \textbf{54.52} & 0.234            & 85.4   & 96.5  \\
              & Ours - FFN(F)         & 2\%      &  IN    & 53.73          & 0.231            & 85.8   & 96.3  \\
              & Ours - Attn(T) FFN(F) & 5\%      &  IN    & \underline{54.14} & \underline{0.238} & 87.1   &  96.9 \\
               & Ours - Attn(T) FFN(F) & 10\%     &  IN    & 54.11          & \textbf{0.240}   & \textbf{88.3}  &  \underline{97.0} \\ 
    \bottomrule
  \end{tabular}
  \label{table:main}
\end{table*}

\subsection{Look-Aside Adapter (LoAA)}
    We propose to use 1D convolution layers as the projection layers within the adapter to facilitate token-wise interactions across time and frequency axes. We call our method Look-Aside Adapter (LoAA). Similar to Equation 1, we define the form of LoAA:\begin{equation}
        h \gets B(h) + f(hU_{\downarrow})U_{\uparrow}
    \end{equation}
    where $U_{\downarrow}$ employs a kernel size of either $1 \times N$ or $N \times 1$, with an input channel of $d$, and an output channel of $r$. Correspondingly, $U_{\uparrow}$ utilizes the same kernel size as $U_{\downarrow}$, with an input channel of $r$, and an output channel of $d$. This means that our adapter module acts as a recognizer of time or frequency patterns, as shown in Figure \ref{fig:interactions}(b). Regarding that a linear layer can be interpreted as $1 \times 1$ convolution layer, we study the effect of kernel size. To ensure a fair comparison, we equalized the number of parameters across each case by adjusting the bottleneck dimension $r$.

    As shown in Table \ref{table:kernel}, LoAA with $1 \times 3$ (Time) and $3 \times 1$ (Frequency) kernel outperforms the parallel adapter with linear layers, denoted as $1 \times 1$ on the table. Considering that an enlargement of kernel size leads to a reduction in bottleneck dimension, this result suggests that a parameter trade-off between the token dimensional axis and the time/frequency axis is beneficial for audio data understanding. Moreover, the low performance in $3 \times 3$ kernel indicates that merely incorporating convolution layers into the adapter, without considering the heterogeneity of audio data, fails to effectively capture audio-specific patterns.

\section{Experiments}


\subsection{Datasets and Tasks}
    We evaluate the Look-Aside Adapter on three commonly used audio and speech benchmarks, including audio classification on EPIC-SOUNDS, Environmental Sound Classification (ESC), and speech classification on Speech Commands.
    \begin{itemize}
        \item \textbf{Pretrain Data (PT-Data)}, including ImageNet \cite{deng09-IN} for vision and AudioSet-2M \cite{gemmeke17-AudioSet} for audio, is used to pretrain baseline models. AudioSet-2M contains approximately 2 million 10-second YouTube clips with weak annotations for 527 types of audio events. 
        \item \textbf{EPIC-SOUNDS (EPIC)} \cite{huh23-EPIC} comprises 78,366 categorized temporal annotations spread across 44 classes, with an average length of 4.9 seconds. Test splits are divided into subsets for audio-based interaction recognition and detection. We employ top-1 accuracy and mean Average Precision (mAP) as evaluation metrics.
        \item \textbf{ESC-50} \cite{piczak15-ESC} comprises 2,000 environmental audio recordings, each lasting 5 seconds, categorized into 50 classes. We evaluate our method using the results from 5-fold cross-validation, using the same data splits as AST.
        \item \textbf{Speech Commands (SPC-2)} \cite{warden18-GSC} is a dataset for two keyword spotting tasks containing 35 speech commands. The training/validation/testing sets consist of 84,843/9,981/11,005 1-second recordings, respectively. 
    \end{itemize}

\subsection{Implementation Details}
\vspace{-0.5em}
    The implementation of our method is based on AST. For the EPIC-SOUNDS dataset that AST has not been tested on, we follow the SSAST setup of EPIC-SOUNDS \cite{huh23-EPIC}, utilizing a learning rate of 1e-4, the AdamW optimizer, and a training duration of 30 epochs without mixup.
    For the ESC-50 dataset, we use a learning rate of 1e-4 for full fine-tuning, the Adam optimizer, and a training duration of 25 epochs without mixup. For PEFT, we choose the learning rate 5e-4 in 5e-5, 1e-4, and 5e-4.
    For the SPC-2 dataset, we use a learning rate of 2.5e-4, the Adam optimizer, and a training duration of 30 epochs with mixup.
    Note that we reshape the flattened tokens into a two-dimensional time-frequency format within our adapter module for the integration of convolution layers. Therefore, we do not overlap patches in our PEFT experiments. We use patch overlapping for full fine-tuning. We conducted our experiments using computing clusters equipped with NVIDIA RTX 3090 GPUs, allocating a single GPU for each experiment.

\subsection{Results}
\vspace{-0.5em}
    We compare the following methods in our experiments:
    \begin{itemize}
    \item \textbf{Image pretrained $\rightarrow$ Audio pretrained}: We introduce two audio models based on AST \cite{Baade2022MAEASTMA, Gong22-SSAST}, which were pretrained on large-scale audio datasets. Following pretraining, these models undergo full fine-tuning for each specific downstream task. The results reported are directly from the papers.
    \item \textbf{Image pretrained Only}: We conduct experiments on audio and speech benchmarks using ImageNet pretrained AST model. We employ both full fine-tuning and linear probing methods to examine the image-based model's ability to learn audio patterns.
    \item \textbf{Image pretrained Only with PEFT (Ours)}: We compare our proposed method with various design choices to parallel adapter. For settings with 2\% trainable parameters, we employ either a time-based LoAA using $1 \times 3$ kernel ($r=36$) with an attention block or a frequency-based LoAA using $3 \times 1$ kernel ($r=36$) with a feed-forward block for simplicity. For 5\% and 10\% parameter settings, we employ a combination of adapters in both blocks, with bottleneck dimensions of $r=36$ and $r=72$, respectively. Parallel adapter parameters are set to 5\%, with a bottleneck dimension of $r=108$.
    \end{itemize}

    The results presented in Table \ref{table:main} offer several key insights. Initially, it is observed that image-pretrained models, through the application of our PEFT method, can achieve performance on par with dedicated audio models. Notably, our approach outperforms SSAST in all examined parameter settings (2\%, 5\%, and 10\%) on the EPIC-SOUNDS dataset and delivers nearly equivalent results on the other two benchmarks, with a performance discrepancy of less than 1\%. 
    
    Furthermore, our method surpasses previous approaches owing to its audio-centric architecture. Our method demonstrates superior performance across all benchmarks when compared to the parallel adapter which has the same number of parameters. This suggests that our proposed method, which enables token-wise interactions across the time and frequency domains, efficiently bridges the gap between audio and image data. 
\vspace{-1em}
\section{Discussions}

\subsection{What is the optimal combination of Look-Aside Adapter modules for a transformer-based model?}

    Recent works \cite{he22-PETL, lin24-peftspeech} empirically showed that parallel insertion is superior to conventional sequential insertion. While those works mix parallel adapters with other PEFT methods, such as LoRA and prefix tuning, to achieve high performance, our method solely uses parallel adapter architecture.
    
    We study the various combinations of time-based LoAA ($1 \times 3$ kernel) and frequency-based LoAA ($3 \times 1$ kernel), using them as parallel adapters. Table \ref{table:block} illustrates the different design choices for attention block (Attn) and FFN block (FFN). When the number of parameters is 5\%, indicating greater parameter efficiency, time-based LoAA for the Attn block and frequency-based LoAA for the FFN block prove to be the best options.

 \begin{table}[t!]
      \caption{Top-1 accuracy of our Look-Aside Adapter Combinations on EPIC-SOUNDS dataset. Best results are highlighted in \textbf{bold}, and second best results are \underline{underlined}.}
      \label{loaa-epic-sounds}
      \centering
      \begin{tabular}{lllll}
      \hline
        \toprule
        \multirow{2}{*}{\textbf{Method}} & \multicolumn{2}{c}{\textbf{\# of params.}} \\ 
            & \multicolumn{1}{c}{\textbf{5\%}}    & \multicolumn{1}{c}{\textbf{10\%}} \\ \hline
         Attn(L) FFN(L)  & 53.40 & 53.85     \\
         Attn(T) FFN(T)  & 53.55 & 54.01     \\
         Attn(F) FFN(F)  & \underline{54.05} & \textbf{54.35}     \\
         Attn(T) FFN(F)  & \textbf{54.14} & \underline{54.11}     \\
         Attn(F) FFN(T)  & 53.90 & 54.04    \\ 
        \bottomrule
      \end{tabular}
      \label{table:block}
    \end{table}
\vspace{-0.5em}
\subsection{What is the effect of fine-tuning by an audio friendly adapter?}
    
    While general PEFT methods should learn domain-specific and task-specific knowledge, our PEFT method should learn the modality difference at the same time. To efficiently transfer knowledge from the different modality, we introduce a new adapter architecture with an inductive bias towards audio data.
    
    Inserting an audio friendly adapter can impact the attention mechanism, potentially leading to smoother attention maps based on the characteristics of the audio Mel-spectrogram.
    An audio friendly adapter can refine the model's focus on relevant parts of the audio Mel-spectrogram by introducing additional parameters that are fine-tuned for downstream tasks. This fine-tuning process can help the model to better distinguish between more and less important features in the frequency-time dimension, leading to smoother and more focused attention distributions.

\subsubsection{Visualization of Attention Map}
\begin{figure}[t]
  \centering
  \includegraphics[width=\linewidth]{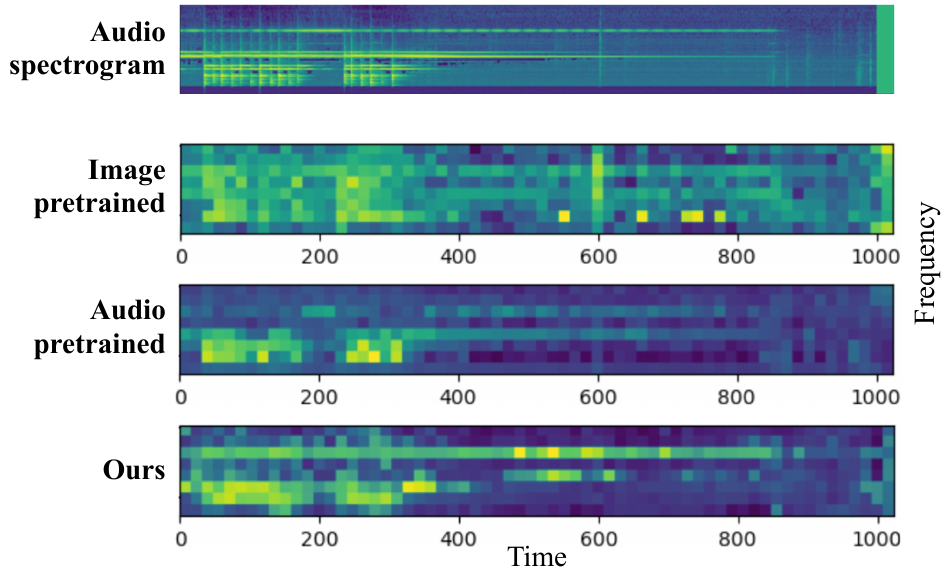}
  \caption{Attention maps of bell sound produced by AST-based models trained in different ways. The audio data is obtained from AudioSet and represented by 128 dimensions of frequency and 1024 timesteps. We examine the self-attention of tokens from the last layer, and display the results for the key timesteps of 320-336. Each token corresponds to $16\times16$ area of the audio spectrogram and the attention maps are made up of $8\times64$ tokens. For clearer visualization, the tokens are upscaled to their original size.}
  \label{fig:attnmap}
\end{figure}

    We visualize the Mel-spectrogram for a bell sound and the attention maps at key timesteps during each stage of training in Figure \ref{fig:attnmap}.
    When utilizing only ImageNet pretrained weights, the attention map is focused on the main parts of the bell sound's Mel-spectrogram but is also noisy.
    
    Addressing this issue, a model with large-scale audio pretraining can better focus on significant aspects of the audio data, thereby reducing overall noise. 
    After large-scale audio pretraining, the resulting attention map has shown a marked reduction in overall noise. It has a tendency to focus on tokens truly crucial for audio classification.
    
    However, instead of relying on extensive audio pretraining, the use of the Look-Aside Adapter (ours) basically reduces overall noise and gives higher attention not only on the essential regions where features of the Mel-spectrogram are prominent but also on the finer details while achieving efficiency.

\section{Conclusion}
    
    The conventional two-stage paradigm, which starts with pretrained vision models, leverages large-scale audio pretraining followed by task-specific fine-tuning to learn the audio-specific knowledge for downstream tasks. However, it is challenging due to the need of substantial audio data and a well-designed learning objective for large-scale audio pretraining. This study proposes bypassing that pretraining stage by directly fine-tuning the vision model with our parallel Look-Aside Adapter (LoAA). It enables efficient knowledge transfer from image models in one step to audio downstream tasks. Our approach outperforms pretrained audio models on EPIC-SOUNDS dataset and also yields comparable results on various other audio and speech datasets. In future work, by delving into the distinctive disparities between image and audio data modalities, we aim to architect a novel multimodal framework with modality-specific encoders.

\bibliographystyle{IEEEtran}
\bibliography{main}

\end{document}